\documentclass[aps, prl, showpacs, twocolumn, amsfonts, amsmath, amssymb, floatfix]{revtex4}

\usepackage[T1]{fontenc}
\usepackage[latin9]{inputenc}
\usepackage{color}
\usepackage{graphicx}
\usepackage{url}
\usepackage[normalem]{ulem}
\usepackage{amsthm}
\usepackage{amsmath}

\def\be{\begin{equation}}
\def\ee{\end{equation}}
\def\bea{\begin{eqnarray}}
\def\eea{\end{eqnarray}}

\begin{document}

\title{Modeling spontaneous breaking of time-translation symmetry}

\author{Krzysztof Sacha} 
\affiliation{
Instytut Fizyki imienia Mariana Smoluchowskiego, 
Uniwersytet Jagiello\'nski, ul. Prof. S. \L{}ojasiewicza 11 PL-30-348 Krak\'ow, Poland}
\affiliation{Mark Kac Complex Systems Research Center, Uniwersytet Jagiello\'nski, ul. Prof. S. \L{}ojasiewicza 11 PL-30-348 Krak\'ow, Poland}


\begin{abstract}
We show that an ultra-cold atomic cloud bouncing on an oscillating mirror can reveal spontaneous breaking of a discrete time translation symmetry. In many-body simulations we illustrate the process of the symmetry breaking that can be induced by atomic losses or by a measurement of particle positions. The results pave the way for understanding and realization of the time crystal idea where crystalline structures form in the time domain due to spontaneous breaking of continuous time translation symmetry.
\end{abstract}

\maketitle

Symmetries of a quantum many-body Hamiltonian are reflected by properties of its eigenstates. However, there are systems whose eigenstates are extremely vulnerable to any symmetry breaking perturbation. Bose gas in a symmetric double well potential, with attractive particle interactions, is a simple example \cite{smerzi97,raghavan99,milburn97,oberthaler07,zin08}. The ground state of the system reflects the symmetry of the external potential but it cannot be easily prepared in an experiment because it is a macroscopic superposition of two Bose-Einstein condensates (BEC) located in different potential wells. Loss of a particle is sufficient to break the symmetry and accumulate all remaining particles in one of the wells. Breaking of a symmetry due to an infinitesimally weak perturbation is called spontaneous symmetry breaking phenomenon.

\begin{figure}
\includegraphics[width=0.95\columnwidth]{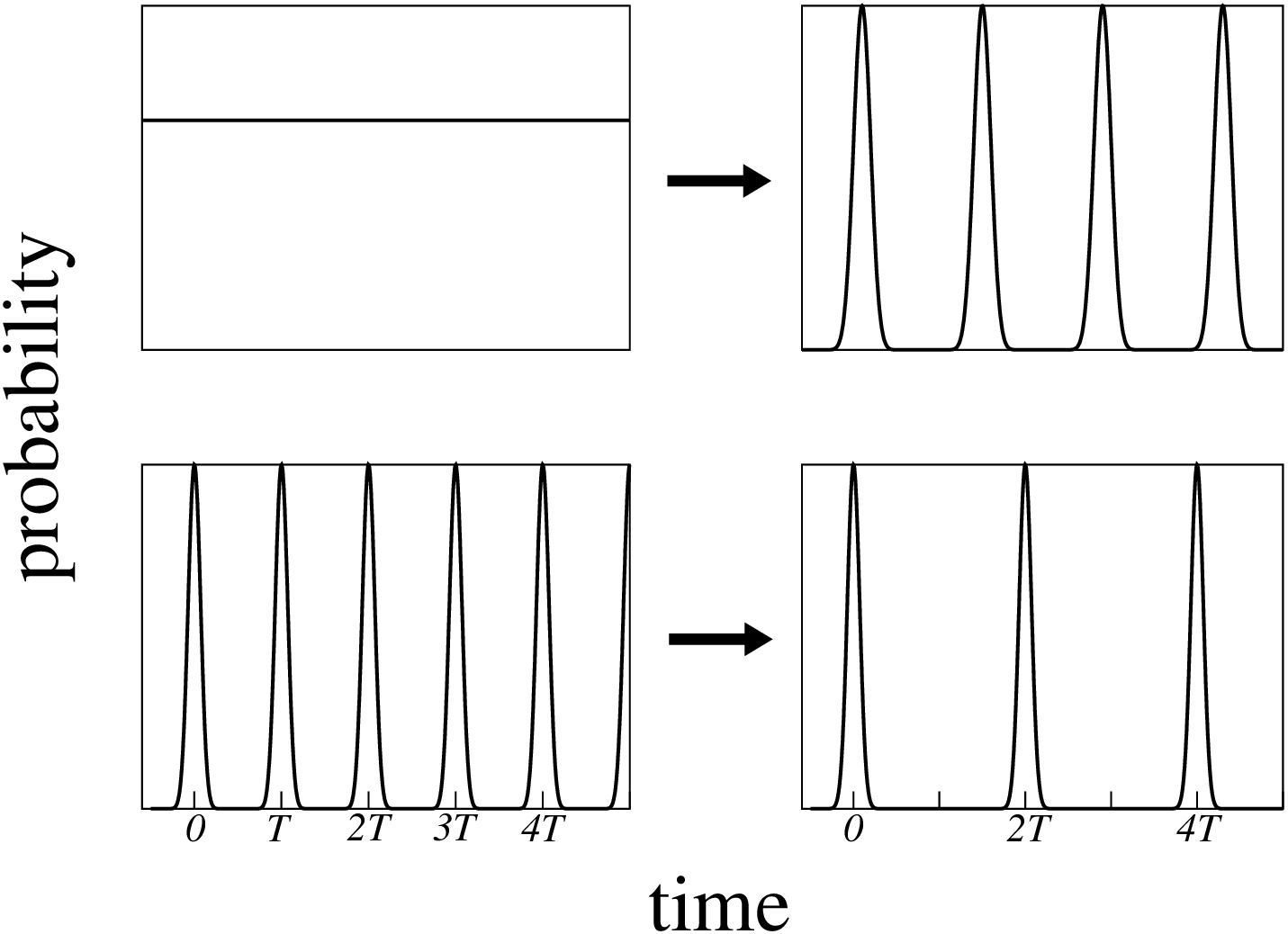}
\caption{Panels show probability of detecting a system at a fixed position versus time. Top panels present effects of spontaneous breaking of continuous time translation symmetry in the course of time crystal formation. Bottom panels show the situation considered in the present paper where spontaneous symmetry breaking results in a change from a discrete time translation symmetry to another discrete one but with a twice longer period $T$.}
\label{schematic}
\end{figure}

Spontaneous breaking of continuous spatial translation symmetry to discrete spatial translation symmetry is responsible for formation of space crystals. Recently it has been proposed that similar phenomenon can also occur in the time domain \cite{wilczek12,li12}. That is, it is possible to invent systems where the ground state of a time-independent Hamiltonian reveals spatially homogeneous flow of particles which under any symmetry breaking perturbation changes spontaneously to periodic motion of spatially inhomogeneous structures. Such a spontaneous breaking of continuous time translation symmetry to a discrete one is termed time crystal formation, see Fig.~\ref{schematic}. Two different systems have been proposed: a bright soliton formed by attractively interacting particles on an Aharonov-Bohm ring \cite{wilczek12} and ions on a ring in the presence of an external magnetic field \cite{li12} (see also \cite{shapere12,chernodub12,wilczek13,mendonca13,yoshii14}). These proposals triggered debate in the literature whether time crystal formation is possible. It seems that different assumptions can lead to contradictory conclusions \cite{zakrzewski12,coleman13,bruno13,wilczek13a,bruno13a,li12a,bruno13b,nozieres13,volovik13,watanabe14}. 

In the present paper we consider a periodically driven many-body system whose description can be reduced to a Hilbert space spanned by two periodically evolving modes. As in the case of a Bose gas in a double well potential \cite{smerzi97,raghavan99,milburn97,oberthaler07,zin08}, a spontaneous symmetry breaking process occurs. In our example, eigenstates of the system possess discrete time translation symmetry which is spontaneously broken to another discrete translation symmetry but with a longer period, see Fig.~\ref{schematic}. The spontaneous symmetry breaking that is predicted within the mean field approach, can be analyzed in full many-body simulations. Moreover it can be realized in ultra-cold atoms experiments. While the system considered in the present paper does not precisely reproduce the idea of time crystals, where spontaneous breaking of the continuous time-translation symmetry is required, it shows a possibility of time-translation symmetry breaking by an infinitesimally small perturbation. The results pave the way for understanding and realization of the time crystal idea.

We consider $N$ atoms that form a Bose-Einstein condensate and bounce on an oscillating, horizontally oriented, atomic mirror in the presence of the gravitational field \cite{buchleitner02,benseghir14,gertjerenken14}. We assume that the atomic cloud is strongly confined in the transverse directions by means of a harmonic potential so that description of the system can be reduced to the one-dimensional Hamiltonian. In the mean field approach all atoms occupy the same single particle wave-function which is a solution of the Gross-Pitaevskii equation (GPE) \cite{pethicksmith}. Thus the GPE of the considered system, in the coordinate frame moving with the mirror \footnote{In order to switch from the laboratory frame to the coordinate frame that moves with the oscillating mirror the following sequence of unitary transformations has been applied: $U_1=e^{ip \frac{\lambda}{\omega}\cos(\omega t)}$, $U_2=e^{iz\lambda\sin(\omega t)}$ and finally $U_3=\exp\left\{i\left[\frac{\lambda}{\omega^2}\sin(\omega t)+\frac{\lambda^2}{8\omega}\sin(2\omega t)-\frac{\lambda^2 t}{4}\right]\right\}$.}, can be written as $\left(H_0+g_0N|\psi|^2\right)\psi=i\partial_t\psi$ with
\be
H_0=-\frac12 \partial_z^2+V(z)+\lambda z\cos(\omega t),
\label{gpe}
\ee
where $V(z)=z$ for $z\ge 0$ and $V(z)=\infty$ for $z<0$.
In (\ref{gpe}) we have used the gravitational units, i.e. $l_0=(\hbar^2/m^2g)^{1/3}$, $t_0=(\hbar/mg^2)^{1/3}$ and $E_0=mgl_0$ for length, time and energy, respectively, where $m$ is atomic mass and $g$ gravitational acceleration. The parameter $\omega$ stands for the frequency of the mirror oscillations, $\lambda/\omega$ is related to the oscillation amplitude and $g_0=2\omega_\perp a(m/\hbar g)^{1/3}$ is the particle interaction strength \cite{pethicksmith} where $a$ is the atomic scattering length and $\omega_\perp$ is the frequency of the harmonic potential along the transverse directions. 

\begin{figure}
\includegraphics[width=0.95\columnwidth]{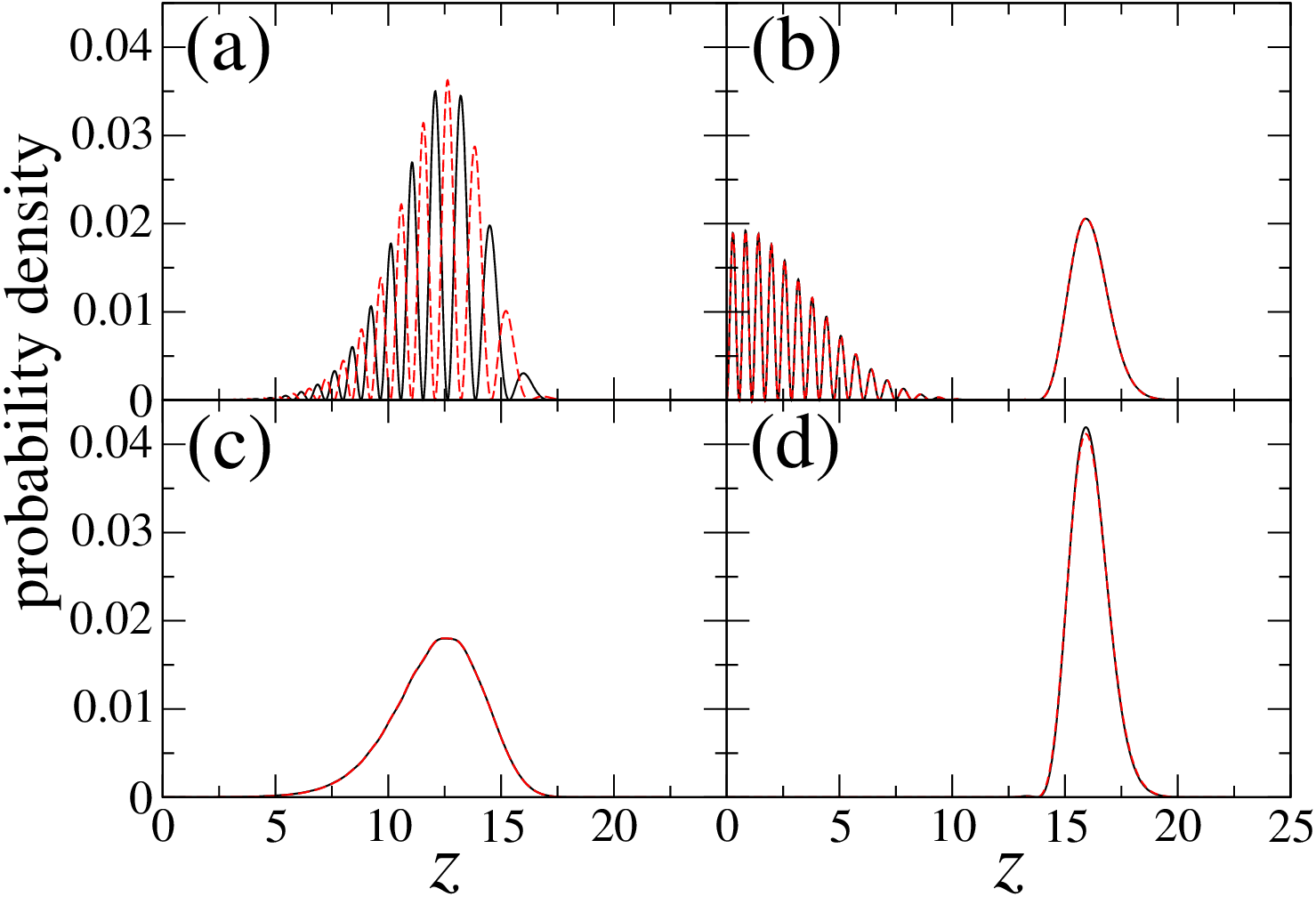}
\caption{(color on line) Panel (a) shows modulus squared of Floquet states $\psi_1(z,t)$ (black solid line) and $\psi_2(z,t)$ (red dash line) of a single particle bouncing on an oscillating mirror in the presence of the gravitational field at $t=0$, see the Hamiltonian (\ref{gpe}), and for $\omega=1.1$ and $\lambda=0.06$. The presented Floquet states are associated with the $s=2$ classical resonance and consist of two wave-packets moving along a periodic orbit. The wave-packets propagate with the period $4\pi/\omega$ and with the $2\pi/\omega$ delay with respect to each other what makes the resulting Floquet states periodic with the period $2\pi/\omega$.  Panel (b) shows the same as (a) but for $t=\pi/\omega$ when the separation of the wave-packets is clearly visible (the interference fringes around $z=0$ are related to the reflection of the wave-packets from the mirror). Bottom panels present a stable solution of the GPE (\ref{nonlinFloq}), for $g_0N=-0.5$, that breaks the original time translation symmetry --- this solution reveals a single wave-packet that propagates along the classical periodic orbit with the period $4\pi/\omega$. Panel (c) is related to $t=0$ while (d) to $t=\pi/\omega$. Black solid lines, in (c) and (d), correspond to the results of the numerical integration of (\ref{nonlinFloq}) while red dash lines (hardly distinguishable from the solid lines) are related to the results of the two-mode approximation. The gravitational units are used, see the text.
}
\label{modes}
\end{figure}
 
Let us begin with the $g_0=0$ case. We, then, deal with a single particle described by the time-dependent Schr\"odinger equation. Periodically driven systems possess quasi-energy eigenstates, so called Floquet states $|\psi_n(t)\rangle$, that are eigenstates of the Floquet operator \cite{buchleitner02}. In our case the Floquet operator reads $H_F=H_0-i\partial_t$. All quasi-energy eigenstates $\psi_n(z,t)$ are time-periodic with the period $2\pi/\omega$. Spectrum of the corresponding eigenvalues $\varepsilon_n$ is not bounded --- it is periodic with the period $\omega$. The Floquet formalism is in full analogy to the Bloch theorem in solid state physics.

Classical motion of a particle bouncing on an oscillating mirror is irregular. If the driving amplitude $\lambda$ is not too big there are regular resonance islands in the phase space that are located around periodic orbits whose periods match $s2\pi/\omega$ where $s$ is integer. If a resonance island is sufficiently large it is possible to find, in the quantum description, a Floquet state that reveals a localized wave-packet moving along a periodic orbit without spreading \cite{buchleitner02}. In the following we concentrate on the $s=2$ resonance. For $s=2$ a single wave-packet moving on the periodic orbit cannot form a Floquet state because the period of the orbit is twice longer than the period of the Floquet states. Consequently the desired Floquet state must be a superposition of two wave-packets propagating along the orbit with the $2\pi/\omega$ delay. Two wave-packets can form two different and orthogonal superpositions. Therefore, we should actually expect two Floquet states which reveal pairs of wave-packets moving along the periodic orbit \cite{buchleitner02}. In Fig.~\ref{modes}(a)-(b) we show an example of such Floquet states for $\lambda=0.06$ and $\omega=1.1$. Two quasi-energy levels associated with the $s=2$ resonance are separated by $\omega/2+J$, where $J$ is a small tunneling rate of individual wave-packets --- if a single wave-packet is prepared initially, it moves on the periodic orbit but after time $1/J$ tunnels to a position of the other missing wave-packet. Semiclassical calculations results in $J\approx \frac{8\sqrt{2}\lambda^{3/4}}{\pi\sqrt{\omega}}\exp(-16\pi\sqrt{\lambda}/\omega^3)$ which is a good estimate provided there is no avoided crossing with any other quasi-energy level \cite{buchleitner02}.

Now let us switch to the $g_0\ne 0$ case. For a periodically driven non-linear system, we may still look for analogues of Floquet states, i.e. time-periodic solutions of the GPE,
\be
\left[H_0+g_0N|\psi(z,t)|^2-i\partial_t\right]\psi(z,t)=\mu\psi(z,t).
\label{nonlinFloq}
\ee
For $g_0N\rightarrow 0$, solutions of (\ref{nonlinFloq}) associated with the classical $s=2$ resonance are identical with the corresponding Floquet states of the linear system. However, when $|g_0N|$ increases, a bifurcation takes place and there appear stable solutions that reveal a single wave-packet moving along the classical periodic orbit, see Fig.~\ref{modes}(c)-(d). The period of such solutions is twice longer than the period of the external driving. Thus, we observe time translation symmetry breaking where stationary mean-field solutions possess different symmetry than the symmetry of the original many-body Hamiltonian. In order to describe the system in the vicinity of the bifurcation point let us apply a two-mode approximation \cite{smerzi97,raghavan99,milburn97,oberthaler07,zin08} which turns out to be very accurate. That is, we look for solutions of (\ref{nonlinFloq}) in the form $\psi\approx \phi_1a_1+\phi_2a_2$ with $\phi_{1,2}=[\psi_1(z,t)\pm e^{-i\omega t/2}\psi_2(z,t)]/\sqrt{2}$, where $\psi_{1,2}$ are the Floquet states of the linear system, presented in Fig.~\ref{modes}(a)-(b), that are normalized according to $\int_0^\infty dz\int_0^{4\pi/\omega}dt|\psi_{1,2}|^2=1$. The $\phi_{1,2}$ modes are single wave-packets propagating along the periodic orbit with the period $4\pi/\omega$. Then, the energy functional of the system can be approximated as follows
\bea
E&=&\int_0^\infty dz\int_0^{4\pi/\omega}dt\;\psi^*\left(H_0-i\partial_t+\frac{g_0N}{2}|\psi|^2\right)\psi \cr
&\approx& -\frac{J}{2}\left(a_1^*a_2+a_2^*a_1\right)+\frac{UN}{2}\left(|a_1|^4+|a_2|^4\right) \cr &&+2U_{12}N|a_1|^2|a_2|^2+{\rm const.},
\label{ef}
\eea
where $J=\varepsilon_2-\varepsilon_1-\omega/2$, $U=g_0\int dzdt|\phi_1|^4$ and $U_{12}=g_0\int dzdt|\phi_1|^2|\phi_2|^2$. Extrema of $E$ are given by solutions of the GPE (\ref{nonlinFloq}). For $\omega=1.1$ and $\lambda=0.06$ we obtain $J=3.6\times10^{-5}$, $U/|g_0|=1.66\times10^{-2}$ and $U_{12}/|g_0|=2.76\times10^{-3}$. If $N|U-2U_{12}|<J$, there are two solutions of the GPE, i.e. $\psi_\pm=(\phi_1\pm\phi_2)/\sqrt{2}$, which are stable. For $g_0<0$ and if $N|U-2U_{12}|>J$, the $\psi_+$ is unstable and there appear two new stable solutions that correspond to the minimal value of $E$, i.e. $\psi_{\pm v}=\sqrt{\frac{1\pm v}{2}}\phi_1+\sqrt{\frac{1\mp v}{2}}\phi_2$, where $v=\sqrt{1-J^2/[N(U-2U_{12})]^2}$. If $g_0>0$, when we cross the bifurcation point, the $\psi_+$ mode remains stable but the $\psi_-$ function, corresponding to the maximal value of $E$, looses its stability and two new stable solutions are born. The two-mode approach predicts accurately the appearance of the bifurcation point and results in a very good approximation for the solutions of the GPE (\ref{nonlinFloq}). In Fig.~\ref{modes}(c)-(d) we present comparison of the solutions obtained within the two-mode approach and by numerical integration of (\ref{nonlinFloq}).  

\begin{figure}
\includegraphics[width=0.95\columnwidth]{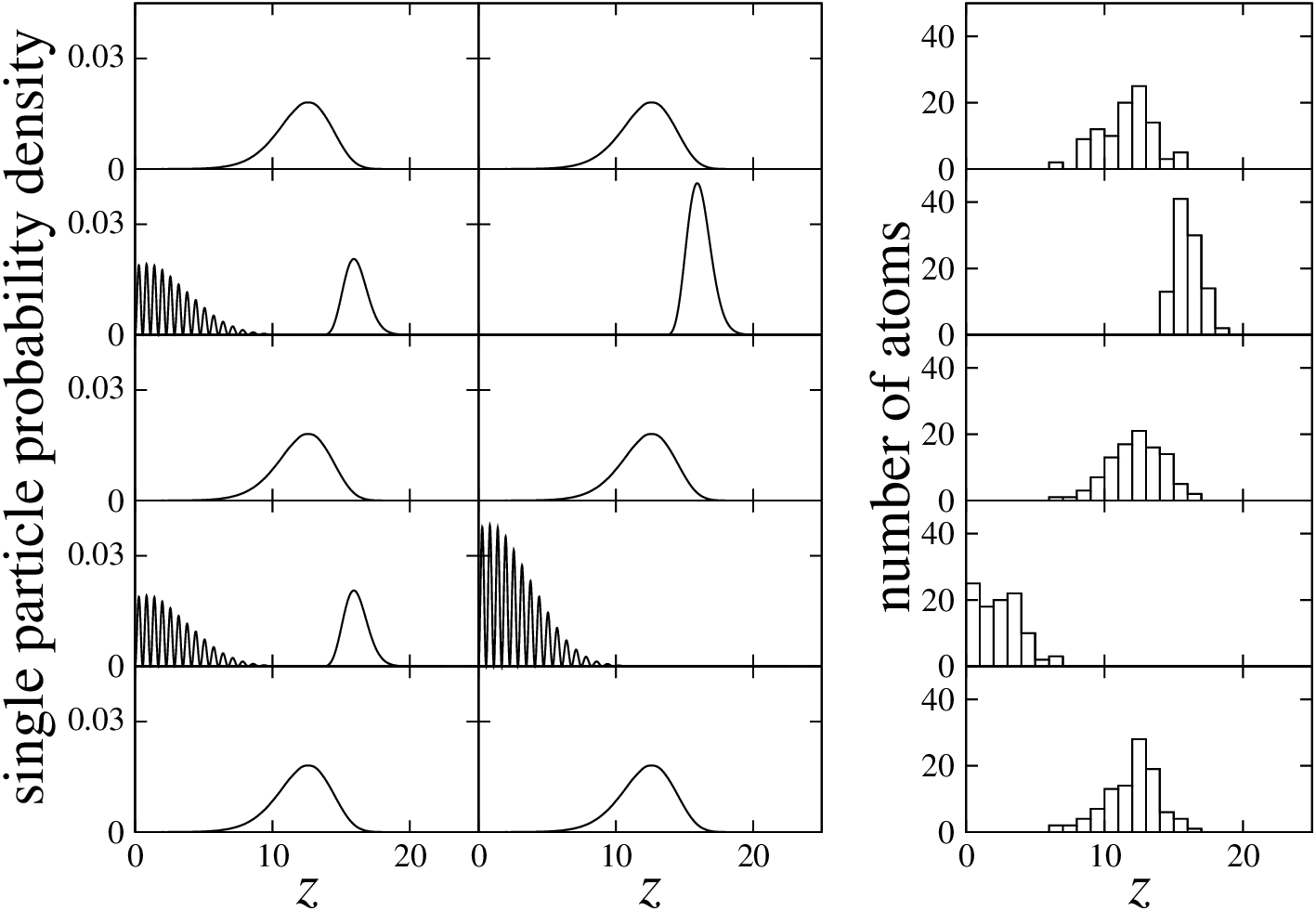}
\caption{Left column shows time evolution of the single particle probability density corresponding to the ground state of the Hamiltonian (\ref{manyH}) for $g_0N=-0.5$, $N=10^4$ and the other parameters as in Fig.~\ref{modes} --- from top to bottom: $t=0$, $\pi/\omega$, $2\pi/\omega$, $3\pi/\omega$ and $4\pi/\omega$. Middle column presents similar data but in the case when every $\pi/\omega$ period positions of 100 atoms are measured. Initially the single particle density $\rho_0(z,0)\approx 0.5N\left[|\phi_1(z,0)|^2+|\phi_2(z,0)|^2\right]$ and because  $|\phi_1(z,0)|^2=|\phi_2(z,0)|^2$, the measurement of particle positions is not able to break the time translation symmetry. When $t$ increases, $|\phi_1(z,t)|^2\ne|\phi_2(z,t)|^2$ and even a measurement of a single particle can break the symmetry and result in a collapse of the system to one of the two propagating wave-packets, see the second panel from the top in the middle column. Right column shows the results of the measurements of positions of 100 atoms, i.e. at $t=0$ we measure positions of 100 atoms, let the remaining atoms evolve and after $\pi/\omega$ we again measure positions of 100 atoms and so on. The histograms presented in the right column indicate that time periodic evolution of the system after the spontaneous time translation symmetry breaking can be observed in a single experimental realization. The gravitational units are used, see the text.
}
\label{singden}
\end{figure}
 
The tunneling splitting $J$ is very small. Thus, even very weak particle interactions lead to the bifurcation and breaking of the original time translation symmetry in the mean-field description. The solutions $\psi_{v}$ and $\psi_{-v}$ correspond to the same extremal value of $E$ and both of them can be realized in an experiment with equal probability --- which of them is realized in a particular experiment is determined in a spontaneous symmetry breaking process. In order to simulate the symmetry breaking process we switch now to the many-body description. We have seen that the two-mode approximation is very accurate in the mean-field approach. Therefore, we apply the same approximation in the many-body problem \cite{milburn97,oberthaler07,zin08}. To this end, the bosonic field operator is truncated, i.e. $\hat\psi(z,t)\approx\phi_1(z,t)\hat a_1+\phi_2(z,t)\hat a_2$, which leads to the many-body Hamiltonian $\hat{\cal H}$ of a similar form as $E$ in (\ref{ef}) but with $a_{1,2}\rightarrow \hat a_{1,2}$. In order to see the symmetry of the many-body Hamiltonian more transparently it is useful to introduce operators $\hat c_{1,2}=(\hat a_1\pm \hat a_2)/\sqrt{2}$ which annihilate particles in the modes $\psi_1(z,t)$ and $e^{-i\omega t/2}\psi_2(z,t)$, respectively, then,
\bea
\hat{\cal H}&\approx& -\frac{J}{2}\left(\hat c_1^\dagger \hat c_1-\hat c_2^\dagger \hat c_2\right)+\frac14 (U-2U_{12})\cr
&&\times \left[(\hat c_1^\dagger)^2\hat c_2^2+(\hat c_2^\dagger)^2\hat c_1^2+2\hat c_1^\dagger \hat c_1\hat c_2^\dagger\hat c_2\right],
\label{manyH}
\eea
modulo constant. From (\ref{manyH}) it is apparent that there are two classes of eigenstates. Eigenstates from the first class are spanned by Fock states with only even occupations of the $\psi_1$ mode while eigenstates from the other class by Fock states with the odd occupations only. It implies that the many-body Floquet states are also eigenstates of the operator that translates the system in time by $2\pi/\omega$. This is illustrated in Fig.~\ref{singden} where we plot time evolution of the single particle density $\rho_0(z,t)=\langle\psi_0|\hat \psi^\dagger(z,t)\hat \psi(z,t)|\psi_0\rangle$ where $|\psi_0\rangle$ is the ground state of the Hamiltonian (\ref{manyH}) corresponding to $g_0N=-0.5$, $N=10^4$ and the same values of $\lambda$ and $\omega$ as in Fig.~\ref{modes}. 

For $g_0<0$, the energy splitting between the ground energy level and the first excited level of (\ref{manyH}) is extremely small \footnote{Note that we measure energy modulo $\omega/2$.} and tends to zero when $N\rightarrow \infty$ but $g_0N=$const --- the Gaussian state approximation in the effective $N$-body approach \cite{zin08} results in the energy splitting proportional to $N\exp(-\alpha N)$ where $\alpha$ is a positive constant. This suggests that for large $N$ even a tiny perturbation can disturb the system dramatically. We will analyze the effect of successive measurements of positions of atoms on the state of remaining particles. Probability density for a measurement of a single atom at position $z$ in the system prepared in a $N$-body state $|\psi\rangle$ is proportional to the single particle density $\langle \psi|\hat \psi^\dagger(z,t)\hat \psi(z,t)|\psi\rangle$. When an atom is localized at a given position $z_1$ in a measurement process at $t=t_1$, it is annihilated from the system, i.e. $|\psi^{(1)}\rangle\propto \hat\psi(z_1,t_1)|\psi\rangle$ is a new state of the remaining $(N-1)$ atoms \cite{javanainen96,dziarmaga06}. The state $|\psi^{(1)}\rangle$ can evolve in time and at $t=t_2$ another measurement of atom position can take place after which there is a novel state of the remaining particles, i.e. $|\psi^{(2)}\rangle\propto \hat\psi(z_2,t_2)|\psi^{(1)}\rangle$ where $z_2$ has been chosen according to the single particle density $\langle \psi^{(1)}|\hat \psi^\dagger(z_2,t_2)\hat \psi(z_2,t_2)|\psi^{(1)}\rangle$. Such a sequential procedure can simulate an intentional measurement process as well as atomic losses that can happen in an atomic system due to, e.g., molecule formation in a three-body collision \cite{dziarmaga03}.

Assume that $N=10^4$ body system is initially prepared in the ground state of the Hamiltonian (\ref{manyH}) $|\psi_0\rangle$ for $g_0N=-0.5$. The single particle density, whose evolution is shown in Fig.~\ref{singden}, reads $\rho_0(z,t)\approx 0.5N\left[|\phi_1(z,t)|^2+|\phi_2(z,t)|^2\right]$. At $t=0$, we have $|\phi_1|^2=|\phi_2|^2$ and the measurements of particle positions are not able to break the time translation symmetry. When $t$ increases the wave-packets $|\phi_1(z,t)|^2$ and $|\phi_2(z,t)|^2$ start moving in the opposite directions and even a single measurement of an atom position can break the symmetry. That is, the initially fragmented BEC \cite{pethicksmith} collapses to a single BEC with all remaining atoms occupying the mode $\phi_1(z,t)$ or $\phi_2(z,t)$ depending on the result of the position measurement. Further evolution of the system and the next measurements of atom positions do not change the form of the state, i.e. once the time translation symmetry is broken, a BEC propagates along the periodic orbit with the period twice longer than the period of the many-body Floquet Hamiltonian. Figure~\ref{singden} illustrates the symmetry breaking process. In this figure we also present the results of the measurements of small fractions of atoms performed every $\pi/\omega$ period which indicates that repeated measurements can reveal the effect of spontaneous symmetry breaking in a single experimental realization.

In order to see signatures of the spontaneous time translation symmetry breaking in an experiment a few elements have to be realized in a laboratory. An oscillating atomic mirror can be created by means of a time modulated evanescent wave \cite{westbrook98}. Initially, in the absence of the mirror oscillations, a BEC has to be prepared in a collectively excited state that will match the resonance with the external driving \cite{buchleitner02} --- in the present paper we have considered the 19th excited state. The excitation can be done by swiping a focus laser beam along the condensate many times \cite{damski01} or by an appropriate phase imprinting \cite{burger99}. Then, the time modulation of the evanescent wave has to be slowly turned on which allows the system to follow adiabatically the desired Floquet state. At the end of this process the system reveals the spontaneous time translation symmetry breaking if particle interactions are attractive, i.e. $g_0<0$, and $J>0$. If they are repulsive, $g_0>0$, no symmetry breaking occurs because the ground state of the Hamiltonian (\ref{manyH}), that is followed adiabatically, is always a nearly perfect BEC and does not suffer from a symmetry breaking perturbation regardless the bifurcation condition is fulfilled or not. Thus, depending on the sign of $g_0$, that can be controlled by means of a Feshbach resonance \cite{pethicksmith}, the final state reveals periodic motion with a period $2\pi/\omega$ or $4\pi/\omega$.

It is worth making comment on the case of strong repulsive particle interactions, i.e. when $NJ\ll(U-2U_{12})$. Then, the ground state of (\ref{manyH}) becomes a Fock state $|N/2,N/2\rangle$ where the same numbers of atoms occupy the $\phi_{1,2}$ modes. At $t$ when $|\phi_1|\approx|\phi_2|$, a measurement of positions of some fraction of atoms reveals interference fringes similar as in Fig.~\ref{modes}(a) but with a random position, cf. interference of two independent BEC's \cite{javanainen96}. If at $t+2\pi/\omega$ another position measurement is performed, the results reveal again interference fringes and again with a random position because the strong interactions push the system again to a fragmented BEC state.

In summary, we have analyzed the process of spontaneous time translation symmetry breaking that is a key element of the formation of time crystals. We have concentrated on a specific system that can be realized experimentally in ultra-cold atomic gases. That is, ultra-cold atoms bouncing on an oscillating mirror in the presence of the gravitational field. We have shown that atomic interactions lead to the time translation symmetry breaking in the mean-field description of the system. Many-body simulations indicate that while the system eigenstates are also eigenstates of the original time translation operator, they are sensitive to infinitesimal symmetry breaking perturbation in the limit of $N\rightarrow\infty$. We have simulated the process of the symmetry breaking that can be induced by measurements of particles positions or by atomic losses.

I thank Kuba Zakrzewski for introducing me to the time crystal phenomenon. I am also grateful to Dominique Delande for the discussion. Support of Polish National Science Centre via project number DEC-2012/04/A/ST2/00088 is acknowledged. The work was performed within the project of Polish-French bilateral programme POLONIUM and the FOCUS action of Faculty of Physics, Astronomy and Applied Computer Science of Jagiellonian University.

\end{document}